\documentclass[final,3p,times,twocolumn]{elsarticle}
\biboptions{comma,sort&compress}
\usepackage{graphicx}
\journal{Nuc. Phys. (Proc. Suppl.)}
\usepackage[english]{babel}
\begin{document}
\begin{frontmatter}
\title{Hadron physics with KLOE--2}
\author{Eryk Czerwi\'nski$^1$ on behalf of KLOE--2 collaboration$^2$}
\address{Laboratori Nazionali di Frascati -- INFN, Via E. Fermi 40, I-00044 Frascati (Rome) Italy}
\begin{abstract}
In the upcoming month the KLOE--2 data taking campaign will start
at the upgraded DA$\Phi$NE $\phi$-factory of INFN Laboratori Nazionali di Frascati.
The main goal is to collect an integrated luminosity of about 20 fb$^{-1}$ in 3-4 years
in order to refine and extend the KLOE program on both kaon physics and hadron spectroscopy.
Here the expected improvements on the results of hadron spectroscopy are presented and briefly discussed.
\end{abstract}
\end{frontmatter}
\footnotetext[1]{{\it Email address:} \texttt eryk.czerwinski@lnf.infn.it}
\footnotetext[2]{KLOE--2 collaboration:
F.~Archilli,
D.~Babusci,
D.~Badoni,
G.~Bencivenni,
C.~Bini,
C.~Bloise,
V.~Bocci,
F.~Bossi,
\mbox{P.~Branchini,}
A.~Budano,
S.~A.~Bulychjev,
P.~Campana,
G.~Capon,
\mbox{F.~Ceradini,}
P.~Ciambrone,
E.~Czerwi\'nski,
E.~Dan\'e,
E.~De~Lucia,
\mbox{G.~De~Robertis,}
A.~De~Santis,
G.~De~Zorzi,
A.~Di~Domenico,
\mbox{C.~Di~Donato,}
B.~Di~Micco,
D.~Domenici,
O.~Erriquez,
G.~Felici,
S.~Fiore,
P.~Franzini,
P.~Gauzzi,
S.~Giovannella,
F.~Gonnella,
E.~Graziani,
F.~Happacher,
B.~H\"oistad,
E.~Iarocci,
M.~Jacewicz,
T.~Johansson,
V.~Kulikov,
A.~Kupsc,
J.~Lee-Franzini,
F.~Loddo,
\mbox{M.~Martemianov,}
M.~Martini,
M.~Matsyuk,
R.~Messi,
S.~Miscetti,
\mbox{D.~Moricciani,}
G.~Morello,
P.~Moskal,
F.~Nguyen,
A.~Passeri,
\mbox{V.~Patera,}
A.~Ranieri,
P.~Santangelo,
I.~Sarra,
M.~Schioppa,
\mbox{B.~Sciascia,}
\mbox{A.~Sciubba,}
M.~Silarski,
S.~Stucci,
C.~Taccini,
L.~Tortora,
\mbox{G.~Venanzoni,}
\mbox{R.~Versaci,}
W.~Wi\'slicki,
M.~Wolke,
J.~Zdebik
}
\section{KLOE--2 detector at upgraded DA$\Phi$NE collider}
\noindent
The KLOE detector setup consists of a large drift chamber (radius from 0.25 to 2.0~m and 3.3~m length) surrounded by an electromagnetic calorimeter.
Both are immersed in 0.52~T axial field of superconducting solenoid~\cite{kloe}. From 2000 to 2006 the KLOE experiment has collected 2.5~fb$^{-1}$ at the peak
of the $\phi$ meson resonance at the $e^{+}e^{-}$ collider DA$\Phi$NE in Frascati plus additional 250~pb$^{-1}$ of off-peak data.

A new beam crossing scheme is operating at DA$\Phi$NE allowing to reduce beam size and increase luminosity (to reach a peak of about 5$\times10^{32}$cm$^{-2}$s$^{-1}$,
a factor of 3 larger than the previously obtained). At the moment, the detector is being upgraded
with small angle tagging devices, to detect both low (Low Energy Tagger - LET) and high (High Energy Tagger - HET) energy $e^{+}e^{-}$ originated from
$e^{+}e^{-}\to e^{+}e^{-}X$ reactions. It is planned to collect around 5~fb$^{-1}$ within one year with this setup.

In a subsequent step a light-material Inner Tracker (IT) will be installed in the region between the beam pipe and the drift chamber to
improve charged vertex reconstruction and to increase the acceptance for low transversal momentum tracks~\cite{it}. Crystal calorimeters (CCALT) will cover
the low $\theta$ region to increase acceptance for very forward electrons and photons down to 8$^\circ$. A new tile calorimeter (QCALT) will be
used for the detection of photons coming from $K_L$ decays in the drift chamber. Implementation of the second step is planned for late 2011.
Further modifications of the DA$\Phi$NE collider, presently under discussion, would allow energies up to 2.5~GeV ($\sqrt{s}$) to be reached without loss of luminosity.
Since the KLOE--2 detector is perfectly suited for data taking also at energies away from the $\phi$ meson mass, a proposal to
perform precision measurements of (multi)hadronic and $\gamma\gamma$ cross sections has also been put forward~\cite{upgradeE}.
The detailed description of the KLOE--2 physics program can be found in Ref.~\cite{kloe2}.
\section{$\gamma\gamma$ physics}
\begin{figure}[h]
  \vspace{-5mm}
  \begin{center}
    \includegraphics[width=0.40\textwidth]{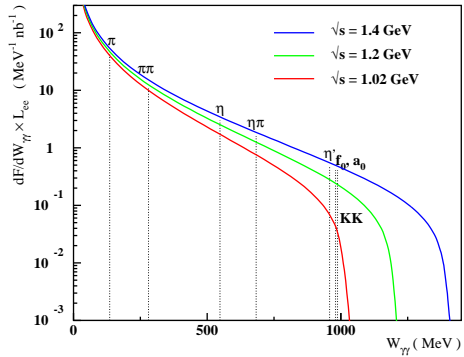}
  \end{center}
  \vspace{-15mm}
  \caption{Differential $\gamma \gamma$ flux function as a function of the center-of-mass energy.}
  \label{flux}
\end{figure}

\noindent
The term $\gamma\gamma$ {\it physics} (or {\it two photon physics}) stands for the study of the reaction:
\begin{equation}
  e^{+}e^{-}\to e^{+}e^{-}\gamma^{*}\gamma^{*}\to e^{+}e^{-}X,
  \label{eq:gg-stat1}
\end{equation}
where $X$ is an arbitrary hadronic state with quantum numbers $J^{PC}=0^{\pm+}, 2^{\pm+} \dots$
and the two photons tend to be quasi-real~\cite{yang}. If no cut is applied to the final-state leptons,
the Weizs\"acker-Williams approximation~\cite{weizsacker} can be used
to understand the main qualitative features of process (\ref{eq:gg-stat1}).
Then the event yield, $N_{eeX}$, can be evaluated according to:
\begin{equation}
  N_{eeX} ~~=~~ L_{ee}\int\frac{{\rm d F}}{{\rm d W_{\gamma\gamma}}}\,
  \sigma_{\gamma\gamma\to X}({\rm W_{\gamma\gamma}})\,{\rm d W_{\gamma\gamma}}~,
  \label{eq:gg-stat2}
\end{equation}
where $W_{\gamma\gamma}$ is the invariant mass of the two quasi--real
photons, $L_{ee}$ is the integrated luminosity, and ${\rm dF}/{\rm dW}_{\gamma\gamma}$ is
the $\gamma\gamma$ flux function:
\begin{equation}
  \frac{{\rm dF}}{{\rm dW}_{\gamma\gamma}}~ ~ = ~~
  \frac{1}{W_{\gamma\gamma}}\ \left(\frac{2\alpha}{\pi}\right)^2\
  \left(\ln\frac{E_b}{m_e}\right)^2 f(z)~,
  \label{eq:gg-stat3}
\end{equation}
where $E_b$ is the beam energy and
\begin{equation}
  f(z) = (z^2+2)^2\ \ln\frac{1}{z}
  -(1-z^2)\ (3+z^2),~z=\frac{\rm W_{\gamma\gamma}}{2E_b}.
  \label{eq:fz}
\end{equation}
Figure~\ref{flux} shows examples of the $\gamma\gamma$ flux functions multiplied by an integrated luminosity $L_{ee}=1$~fb$^{-1}$, as
a function of the $\gamma\gamma$ invariant mass for different center-of-mass energies;
threshold openings of different hadronic states are indicated.
Previous experiments measured the $\gamma\gamma$ cross section for pseudoscalar meson production in the range
\mbox{$7<\sqrt{s}<35$~GeV}. A low energy $e^{+}e^{-}$ collider, such as DA$\Phi$NE, compensates the small cross section value
with the high luminosity.
\section{Meson transition form factors}
\noindent
The transition form factors
$\mathcal{F}_{P\gamma^*\gamma^*}(m^2_P,q_1^2,q_2^2)$ at space-like
momentum transfers can be measured with process~(\ref{eq:gg-stat1}).
They are important to discriminate among different phenomenological
models relevant for the hadronic light-by-light scattering contribution to the $g-2$ of the
muon~\cite{Jegerlehner:2009ry}.

The form factor at negative $q^2$ appears in the production cross
section of $\pi^0$, $\eta$ and $\eta^{\prime}$ mesons in the reaction $e^+ e^- \to e^+ e^- P$.
By detecting one electron at large angle with respect to the beams,
the form factor $\mathcal{F}_{P\gamma\gamma^*}(m^2_P,Q^2,0)$ with one quasi--real and one
virtual space-like photon ($Q^2=-q^2$) can be measured for the on--shell pseudoscalar meson.
For both $\pi^0$ and $\eta$ mesons the region below 1~GeV$^2$ is still poorly known but can be covered by KLOE--2.
Furthermore, by selecting events in which both $e^+$ and $e^-$
are detected by the drift chamber (instead of the tagger devices)
\mbox{KLOE--2} can provide experimental information on form factors
$\mathcal{F}_{P\gamma^* \gamma^*}(m^2_P,Q_1^2,Q^2_2)$, with two
virtual photons.
\section{$\gamma\gamma\to\pi\pi$}
\noindent
The two photon production of hadronic resonances is often advertised as one of
the clearest ways of revealing their
composition~
\cite{vanBeveren:2008st,Hanhart:2007wa,Branz:2008ha,Volkov:2009pc,Mennessier:2008kk,Achasov:2007qr,Giacosa:2008xp,Giacosa:2007bs,Klempt:2007cp,Pennington:2007zy,Barnes:1985cy}.
KLOE--2 with the study of $e^+ e^- \to e^+ e^- \pi \pi$ decays can improve the experimental precision in the following energy ranges
contributing to the solution of the open questions on low--energy hadron physics:
\begin{itemize}
\item
280-450 MeV:
The Mark II experiment~\cite{Boyer:1990vu} is the only one that has made a special measurement of the normalized
cross-section for the $\pi^+\pi^-$ channel near threshold, however,  their data have very large error-bars;
\item
450-850 MeV:
Measure $\pi\pi$ production in this region in both charge
modes for our understanding of strong QCD coupling and the nature of the vacuum~\cite{Pennington:2007yt};
\item
850-1100 MeV:
Accurate measurement of the $\pi^+\pi^-$ and
$\pi^0\pi^0$ cross-sections (integrated and differential).
\end{itemize}
The $e^+ e^- \to e^+ e^- \pi \pi$ process is a clean electromagnetic probe to investigate
the nature of the $\sigma$ meson through the analysis of the
$\pi\pi$ invariant mass which is expected
to be plainly affected by the presence of this scalar meson.
A precision measurement of the cross-section of $\gamma\gamma\to\pi^+\pi^-$
and $\gamma\gamma\to\pi^0\pi^0$ would also complete the information from
previous experiments allowing the determination of the $\gamma\gamma$ couplings
of the scalar mesons.
\subsection{$\gamma\gamma\to\pi^{0}\pi^{0}$}
\begin{figure}[h] 
  \begin{center}
    \includegraphics[width=0.40\textwidth]{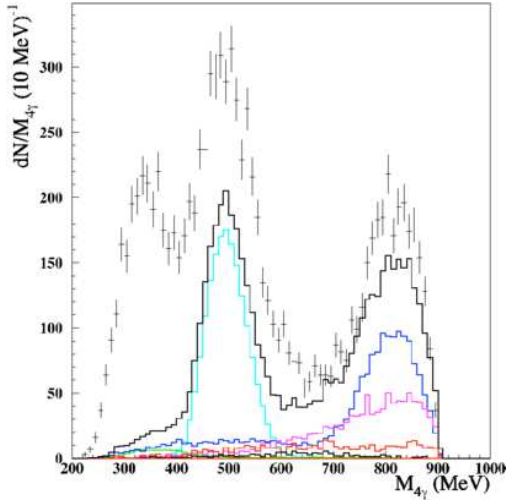}
  \end{center}
  \vspace{-6mm}
  \caption{Preliminary spectrum of the 4 photons invariant mass obtained with KLOE detector, compared with sum of the expected
  backgrounds. Peak of the $K_{S}\to\pi^{0}\pi^{0}$ decay and structures related to other processes with two
  $\pi^{0}$ are visible: $\omega(\to\pi^{0}\gamma)\pi^{0}$ and $f_{0}(980)(\to2\pi^{0})\gamma$.
  The cut on $M_{4\gamma}<900$~MeV is explained by the requirement on the total energy in the calorimeter.}
  \label{4photon} 
\end{figure}

\noindent
The interest in this process is given by the $\sigma\to\pi\pi$ contribution~\cite{pdg}.
The determination of the $\sigma\gamma\gamma$ coupling can be compared with that of pseudoscalars or other
scalar states to clarify their quark structure.
From Figure~\ref{4photon}, an excess of about 4000 events with respect to the expected background
is evident at low 4 photons invariant mass ($M_{4\gamma}$) values, consistent in shape with expectations~\cite{federico}
from $e^{+}e^{-}\to e^{+}e^{-}\pi^{0}\pi^{0}$ events. The precise yield estimate depends
on assumptions for the background processes. Systematic study of the differential cross
section and understanding of the $\sigma\to\pi\pi$ contribution are in progress.

The studies point out that KLOE-2 with an integrated luminosity
at the $\phi$ peak of $L=$ 5 fb$^{-1}$ can measure
the $\gamma\gamma\to\pi^0\pi^0$ cross-section with the same energy
binning obtained from Crystal Ball \cite{Marsiske:1990hx},
reducing the statistical uncertainty in each bin to 2\%.
\section{$\eta\to\pi^{0}\gamma\gamma$}
\begin{figure}[h]
  \vspace{-5mm}
  \begin{center}
    \includegraphics[width=0.40\textwidth]{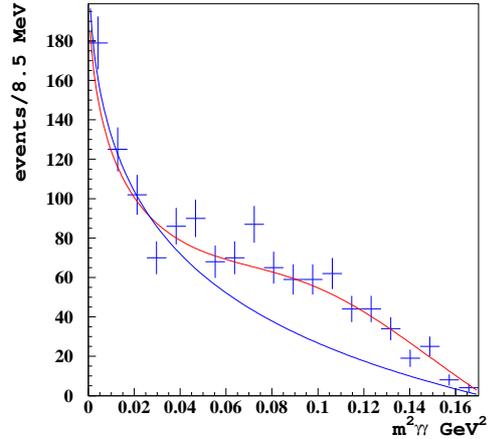}
  \end{center}
  \vspace{-7mm}
  \caption{The $m^2_{\gamma \gamma}$ distribution in
    $\eta \to \pi^0 \gamma \gamma$ decays expected at KLOE--2 for the
    VMD+$a_0(980)$ model, with constructive and destructive
    interference term.  Crosses are the simulated experimental
    data assuming 5\% constant efficiency as a function of
    $m^2_{\gamma \gamma}$ and constructive interference.}
  \label{fig:experiments2}
\end{figure}

\noindent
Using the KLOE preliminary result on the branching fraction and the
analysis efficiency obtained of $\sim5\%$, 1300 $\eta \to \pi^0 \gamma \gamma$
events are expected from the first year of data-taking
at KLOE--2, thus allowing an accuracy of 3\% to be reached on the BR($\eta\to\pi^{0}\gamma\gamma$.  Moreover, KLOE--2 can provide
the $m_{\gamma \gamma}$ distribution with sufficient precision to solve the ambiguity connected to the sign
of the interference between VMD and scalar terms as shown in Figure~\ref{fig:experiments2}.
\section{$\eta'\to\eta\pi\pi$}
\begin{figure}[h] 
  \vspace{-4mm}
  \begin{center}
    \includegraphics[width=0.40\textwidth]{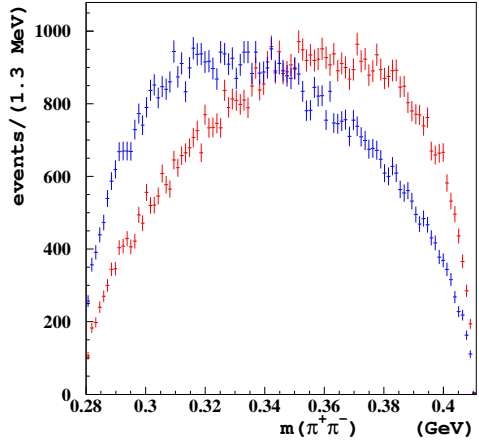}
  \end{center}
  \vspace{-5mm}
  \caption{The $m_{\pi^+ \pi^-}$ distribution
   in the $\eta' \to \eta \pi^+ \pi^-$ decay with the $\sigma$ meson
   (right--centered distribution) and without (left--centered distribution) contribution.}
  \label{2pi0} 
\end{figure}

\noindent
In the $\eta'\to \eta \pi^+ \pi^- $ and $\eta' \to \eta \pi^0 \pi^0 $,
the $\pi\pi$ system is produced mostly with scalar quantum
numbers. Indeed, the available kinetic energy of the $\pi^+ \pi^-$
pair is \mbox{[0, 137] MeV,} suppressing high angular momentum contribution.
Furthermore, the exchange of vector mesons is forbidden by G-parity
conservation. For these reasons, only scalar mesons can participate to
the scattering amplitude. The decay can be mediated by the $\sigma$,
$a_0(980)$ and $f_0(980)$ exchange and by a direct contact term due to
the chiral anomaly \cite{Fariborz:1999gr}.
The scalar contribution can be determined by fitting the Dalitz plot of
the $\eta' \to \eta \pi \pi $ system. The golden channel for KLOE--2 is the
decay chain $\eta' \to \eta \pi^+ \pi^- $, with $\eta \to \gamma \gamma$.
The signal can be easily identified from the $\eta$ and $\eta'$
invariant masses. Such final state was already studied at KLOE to measure
the branching fraction of the $\phi \to \eta' \gamma$ decay
\cite{Aloisio:2002vm}. The analysis efficiency was 22.8\%, with 10\%
residual background contamination.
With ${\mathcal O}(10)$ fb$^{-1}$, we expect
80,000 fully reconstructed events.
In Figure~\ref{2pi0} the $m_{\pi^+ \pi^-}$ invariant mass
distribution is shown with and without the $\sigma$ contribution with the
expected KLOE--2 statistics.
\section{$\eta/\eta'$ mixing}
\noindent
The $\eta'$ meson, being almost a pure SU(3)$_{\small \rm Flavor}$
singlet, is considered a good candidate to host a gluon condensate.
KLOE has extracted the $\eta'$ gluonium content
and the $\eta$-$\eta'$ mixing angle~\cite{Ambrosino:2009sc}
according to the model of Ref.~\cite{Rosner:1982ey}.
The $\eta$ and $\eta'$ wave functions can be decomposed in three terms:
the $u,d$ quark wave function
$\left | q \bar{q} \right > = \frac{1}{\sqrt{2}} 
\left(\left | u\bar{u} \right >+\left |d\bar{d} \right >\right)$,
the $\left | s \bar{s}\right >$ component and the gluonium
$\left| GG \right>$.  The wave functions are written as:
$\left |\eta'\right >  =  \mathrm{cos}\psi_G\,\mathrm{sin}\psi_P \left |q \bar{q} \right > +
  \mathrm{cos}\psi_G\,\mathrm{cos}\psi_P \left |s \bar{s} \right > +
  \mathrm{sin}\psi_G \left |GG \right> \rm{~and~} \left |\eta \hphantom{'} \right>  = 
  \mathrm{cos}\psi_P \left |q \bar{q} \right >  - 
  \mathrm{sin}\psi_P \left |s \bar{s}  \right >$
where $\psi_P$ is the $\eta$-$\eta'$ mixing angle and
$Z^2_G = \mathrm{sin}^2 \psi_G$ is the gluonium fraction in the
$\eta'$ meson. The $Z^2_G$ parameter can be interpreted as the mixing
with a pseudoscalar glueball.

\begin{figure}[h] 
  \vspace{-4mm}
  \begin{center}
    \includegraphics[width=0.40\textwidth]{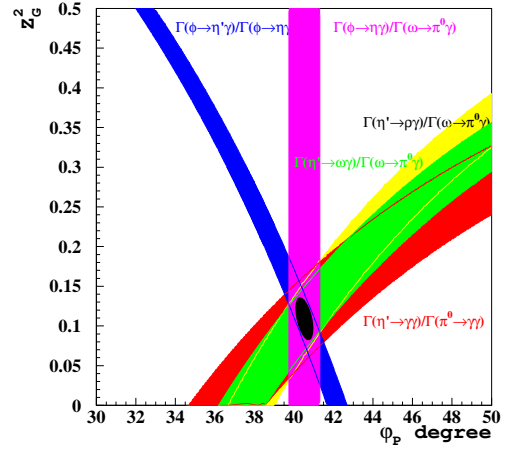}
    \includegraphics[width=0.40\textwidth]{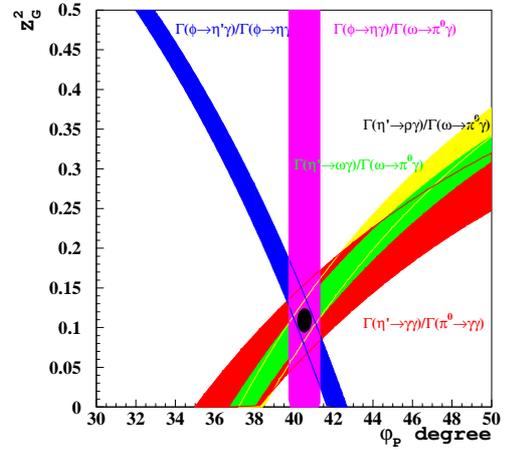}
  \end{center}
  \vspace{-5mm}
  \caption{The 68\% C.L. region in the $\psi_P,Z^2_G$ plane.  Top: only the $\eta^\prime$ branching ratios are
    improved to 1\% precision.  Bottom: the $\eta'$ total width is also lowered to 1.4\%.}
  \label{mixing} 
\end{figure}

With the KLOE--2 data-taking above the $\phi$ peak, e.g.,
at  $\sqrt{s} \sim 1.2$ GeV,
it will be possible to measure the $\eta'$
decay width $\Gamma(\eta' \to \gamma \gamma)$ through the measurement of the reaction
$\sigma(e^+ e^- \to e^+ e^- (\gamma^* \gamma^*) \to e^+ e^- \eta')$.
The measurement to 1\% level of both the cross section and the
$BR(\eta' \to \gamma \gamma)$
would bring the fractional error on the
$\eta'$ total width,
$\Gamma_{\eta'} = \Gamma(\eta' \to \gamma \gamma) /
BR(\eta' \to \gamma \gamma)$,
to $ \sim$1.4\%.
Figure~\ref{mixing} shows the 68\% C.L. region in
the $\psi_P,Z^2_G$ plane obtained with the improvements discussed in this section.
The comparison of the top to bottom panels makes evident how the fit
accuracy increases
with the  precision measurement of the $\eta^\prime$ total width.

\end{document}